\documentclass[pdf,aps,preprint,superscriptaddress]{revtex4}

\usepackage{graphicx}

\begin{document}

\title{ForMAX - a beamline for multiscale and multimodal structural characterization of hierarchical materials}

\author{K. Nyg{\aa}rd}
\email{kim.nygard@maxiv.lu.se}
\affiliation{MAX IV Laboratory, Lund University, Lund, Sweden}

\author{S.~A. McDonald}
\author{J.~B. Gonz{\'a}lez}
\author{V. Haghighat}
\affiliation{MAX IV Laboratory, Lund University, Lund, Sweden}

\author{C. Appel}
\affiliation{MAX IV Laboratory, Lund University, Lund, Sweden}
\affiliation{Paul Scherrer Institut, Villigen PSI, Switzerland}

\author{E. Larsson}
\affiliation{MAX IV Laboratory, Lund University, Lund, Sweden}
\affiliation{Division of Solid Mechanics, Lund University, Lund, Sweden}

\author{R. Ghanbari}
\affiliation{MAX IV Laboratory, Lund University, Lund, Sweden}
\affiliation{Department of Industrial and Materials Science, Chalmers University of Technology, Gothenburg, Sweden}

\author{M. Viljanen}
\author{J. Silva}
\author{S. Malki}
\author{Y. Li}
\author{V. Silva}
\author{C. Weninger}
\author{F. Engelmann}
\author{T. Jeppsson}
\author{G. Felcsuti}
\affiliation{MAX IV Laboratory, Lund University, Lund, Sweden}

\author{T. Ros{\'e}n}
\affiliation{Department of Fibre and Polymer Technology, Royal Institute of Technology, Stockholm, Sweden}
\affiliation{Wallenberg Wood Science Center (WWSC), Royal Institute of Technology, Stockholm, Sweden}

\author{K. Gordeyeva}
\affiliation{Department of Fibre and Polymer Technology, Royal Institute of Technology, Stockholm, Sweden}

\author{L.~D. S{\"o}derberg}
\affiliation{Department of Fibre and Polymer Technology, Royal Institute of Technology, Stockholm, Sweden}
\affiliation{Wallenberg Wood Science Center (WWSC), Royal Institute of Technology, Stockholm, Sweden}

\author{H. Dierks}
\author{Y. Zhang}
\author{Z. Yao}
\author{R. Yang}
\author{E.~M. Asimakopoulou}
\author{J.~K. Rogalinski}
\author{J. Wallentin}
\author{P. Villanueva-Perez}
\affiliation{Synchrotron Radiation Research, Lund University, Lund, Sweden}

\author{R. Kr{\"u}ger}
\affiliation{Medical Radiation Physics, Lund University, Lund, Sweden}

\author{T. Dreier}
\affiliation{Medical Radiation Physics, Lund University, Lund, Sweden}
\affiliation{Excillum AB, Kista, Sweden}

\author{M. Bech}
\affiliation{Medical Radiation Physics, Lund University, Lund, Sweden}

\author{M. Liebi}
\affiliation{Paul Scherrer Institut, Villigen PSI, Switzerland}
\affiliation{Institute of Materials, {\'E}cole Polytechnique F{\'e}d{\'e}rale de Lausanne (EPFL), Lausanne, Switzerland}
\affiliation{Department of Physics, Chalmers University of Technology, Gothenburg, Sweden}

\author{M. Bek}
\affiliation{Department of Industrial and Materials Science, Chalmers University of Technology, Gothenburg, Sweden}
\affiliation{FibRe-Centre for Lignocellulose-based Thermoplastics, Department of Chemistry and Chemical Engineering, 
Chalmers University of Technology, Gothenburg, Sweden}

\author{R. K{\'a}d{\'a}r}
\affiliation{MAX IV Laboratory, Lund University, Lund, Sweden}
\affiliation{Department of Industrial and Materials Science, Chalmers University of Technology, Gothenburg, Sweden}
\affiliation{FibRe-Centre for Lignocellulose-based Thermoplastics, Department of Chemistry and Chemical Engineering, 
Chalmers University of Technology, Gothenburg, Sweden}
\affiliation{Wallenberg Wood Science Center (WWSC), Chalmers University of Technology, Gothenburg, Sweden}

\author{A.~E. Terry}
\author{H. Tarawneh}
\author{P. Ilinski}
\author{J. Malmqvist}
\author{Y. Cerenius}
\affiliation{MAX IV Laboratory, Lund University, Lund, Sweden}

\begin{abstract}
The ForMAX beamline at the MAX IV Laboratory provides multiscale and multimodal structural characterization of 
hierarchical materials in the nm to mm range by combining small- and wide-angle x-ray scattering with full-field 
microtomography. The modular design of the beamline is optimized for easy switching between different experimental 
modalities. The beamline has a special focus on the development of novel, fibrous materials from forest resources, but 
it is also well suited for studies within, e.g., food science and biomedical research.   
\end{abstract}

\maketitle                        

\section{Introduction}

Many natural and synthetic materials are hierarchical, exhibiting important structure at several different length scales 
that govern the material's properties~\cite{lakes93,fratzl07,gibson12}. Wood is an archetypical example, with the 
assembly of the load-bearing cellulose at nano-, micro-, and macroscopic scales determining its mechanical properties. 
In order to understand the structure-function relationship in such materials, we need access to multiscale structural 
characterization. Moreover, we need sufficient temporal resolution to allow monitoring how the structure evolves 
\emph{in situ} or \emph{in operando} during external stimuli or processing of the material. 

The ForMAX beamline of MAX IV addresses this need for structural characterization of hierarchical materials. A key 
feature is its modular design, that allows temporally resolved multiscale structural characterization of bulk materials 
owing to easy and fast switching between complementary experimental modalities: small- and wide-angle x-ray scattering 
(SWAXS) in the nm regime~\cite{glatter82,pauw13} and full-field synchrotron x-ray microtomography (SR\(\mu\)CT) in the \(\mu\)m to mm 
regime~\cite{maire14}. Both of these techniques are applicable to a wide range of materials and suitable for temporally 
resolved experiments. We foresee that SWAXS will often be carried out in scanning imaging mode using a focused x-ray 
beam, either as SWAXS-based microscopy~\cite{lichtenegger99,bunk09}, tomography~\cite{feldkamp09,jensen11}, or 
tensor tomography~\cite{liebi15,schaff15}, covering seven orders of magnitude in length scales and hence being 
particularly useful for structural characterization of hierarchical materials.  

ForMAX is externally funded, with the objective to support research on new materials from renewable forest resources. 
Its construction was funded by the Knut and Alice Wallenberg foundation (kaw.wallenberg.org), while the operation costs for ten years are  
covered by Swedish industry via Treesearch (www.treesearch.se) - a national research platform for the development of 
new materials and specialty chemicals from the forest. Access for both Treesearch members and general users is granted 
through common calls for proposals, with half of the user beam time reserved for academic and industrial members of 
Treesearch. For a brief background to ForMAX, see~\cite{mcentee23}.

In the following we outline the technical design of the beamline, data acquisition, and data processing, with a focus 
on the needs of the users. We conclude by providing a first benchmarking of the beamline and a few examples of 
multiscale and multimodal structural characterization available at ForMAX. 

\section{Technical design}
\begin{figure}
\includegraphics[width=15cm]{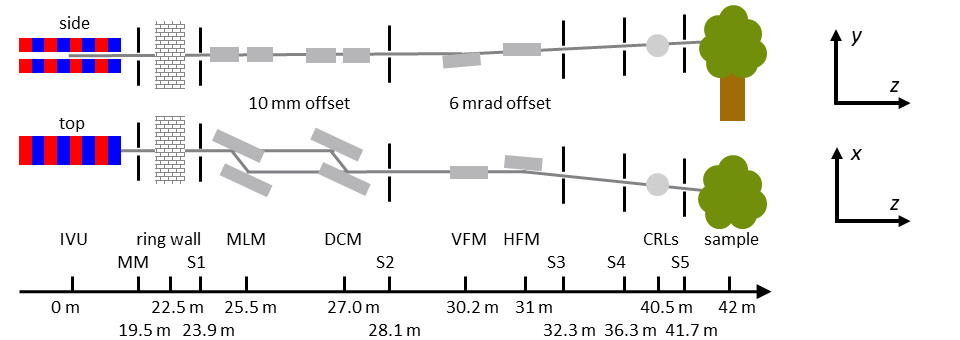}
\caption{Schematic side and top views of the beamline optics and beam-conditioning components along the beamline (not to scale), 
with approximate distance from the source at the bottom. Diagnostic components are omitted for clarity. The MAX IV coordinate system 
is depicted to the right. 
IVU: in-vacuum undulator; 
MLM: multilayer monochromator; 
DCM: double crystal monochromator; 
VFM: vertically focusing mirror; 
HFM: horizontally focusing mirror; 
CRLs: compound refractive lenses; 
MM: front end movable mask;
S1: white-beam slits; 
S2-S5: monochromatic slits. 
}
\label{FIG:BL} 
\end{figure}

The combination of SWAXS and SR\(\mu\)CT provides a number of technical challenges, in particular when 
applied \emph{in situ} or \emph{in operando} to fibrous materials such as wood-based materials:

\begin{itemize}

\item In SR\(\mu\)CT one monitors the attenuated beam directly downstream of the sample (i.e., in forward scattering 
direction), while in SAXS one collects scattering data further downstream at small angles. Since the SR\(\mu\)CT full-field 
microscope blocks the view of the SAXS detector, we have devised 
a strategy for easy movement of the former in and out of the x-ray beam. 

\item WAXS from fibrous materials exhibits anisotropy, reflecting the orientation of the crystalline fibers, fibrils, 
or filaments. When mapping out nanoscale orientation in such materials, one needs to be able to collect WAXS data in 
all directions of the scattering plane~\cite{lichtenegger99}. In order to facilitate scanning SWAXS imaging experiments 
on these materials, we have therefore chosen a custom WAXS detector with a hole in the center, that passes through 
the SAXS signal while simultaneously catching anisotropic WAXS data. 

\item SWAXS is often carried out at moderate x-ray energies \(E \approx 10\)~keV in order to reach small scattering 
vector moduli \(q\), while it is advantageous to carry out SR\(\mu\)CT at higher x-ray energies 
(\(\geq 20\)~keV) for enhanced phase contrast. As a compromise we operate ForMAX in the 8-25 keV energy range, 
which is particularly suitable for soft materials. 

\item Whereas the small divergence of the x-ray beam at the MAX IV 3 GeV storage ring~\cite{tavares18} is beneficial 
for SWAXS experiments, it limits the natural beam size at the sample position in full-field imaging. As a compromise, 
we have placed the sample relatively far downstream of the source (42 m from source), while still allowing a reasonable 
sample-to-detector distance for SAXS experiments. Moreover, we will install secondary beam-expanding optics in the 
experimental station to facilitate full-field imaging. 
 
\item In order to obtain a clean x-ray beam for SAXS experiments, we need to reject higher harmonics of the monochromator 
by passing the beam via x-ray mirrors. In the full-field imaging mode, the slope errors of the mirrors cause parasitic 
striation of the x-ray beam. We mitigate the effect of striation by shape compensation of the mirrors.   

\item Due to the high photon density at fourth generation sources like the MAX IV 3 GeV storage ring~\cite{tavares18}, 
radiation damage in organic samples is a major issue that we need to assess and mitigate case by case. This also holds true for 
full-field imaging, that has traditionally been less prone to beam-induced radiation damage due to a large beam size.   

\item Finally, in order to accomodate various sample environments, such as a rheometer or a mechanical load device 
with controlled atmosphere, we need an experimental table that is spacious and has a relatively large load capacity.  

\end{itemize}
In Table~\ref{TAB:MAXIV} we summarize the main parameters of the MAX IV 3 GeV storage ring, while in Table~\ref{TAB:BL} 
and Fig.~\ref{FIG:BL} we present the main components of the ForMAX beamline. 

Throughout this article, we employ MAX IV's coordinate system: the lateral \emph{x} axis with positive direction 
outbound from the ring, the vertical \emph{y} axis with positive direction upwards, and the longitudinal \emph{z} 
axis with positive direction downstream from the source. The positive direction of each rotation around the Cartesian 
axes (\emph{Rx}, \emph{Ry}, and \emph{Rz}) is given by the right-hand rule.

\subsection{Undulator and front end}

\begin{figure}
\includegraphics[width=10cm]{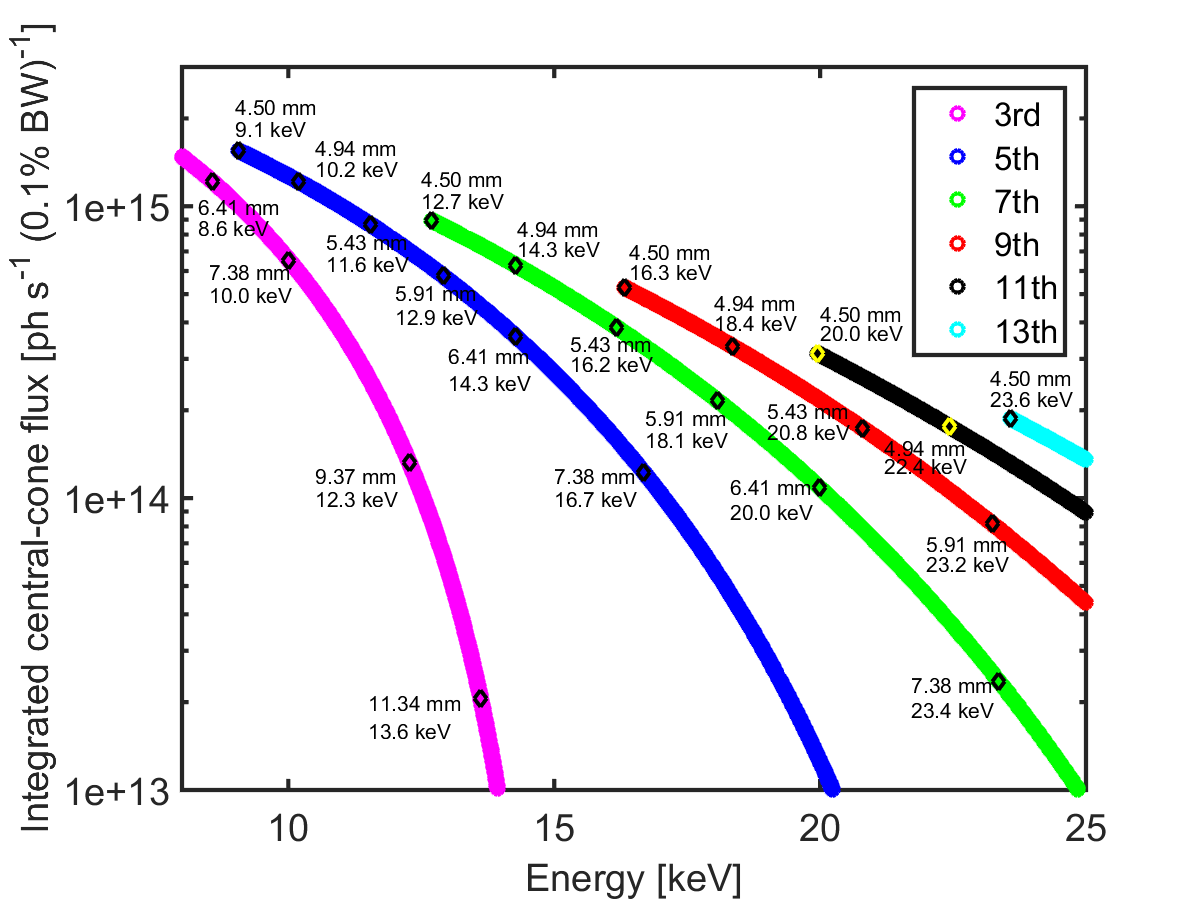}
\caption{Approximate integrated central-cone flux versus x-ray energy \cite{kim09}, shown for odd undulator harmonics. 
Selected insertion device gaps are specified for convenience.}
\label{FIG:ID} 
\end{figure}

ForMAX is equipped with a 3~m long room-temperature, in-vacuum undulator from Hitachi Metals. The maximum 
effective deflection parameter is \(K = 1.89\) at the minimum magnetic gap of 4.5 mm and the measured phase error is 
within specification for all operational gaps. In order to cover the energy range of 8-25 keV, we make use of the fifth to 
thirteenth harmonics of the undulator as shown in Fig.~\ref{FIG:ID}. Similar to other beamlines around the MAX IV 3 
GeV storage ring~\cite{ursby20,johansson21}, the undulator exhibits narrow harmonic peaks, \(\Delta E <100\)~eV 
(full width at half maximum; FWHM). We summarize the main parameters of the undulator in Table~\ref{TAB:ID}.

The front end serves as the interface between the MAX IV 3 GeV storage ring and the ForMAX beamline and was provided by 
Toyama. It is part of 
personal and machine safety systems; it ensures safe access to the optical hutch and safe equipment operation. It 
includes safety and photon shutters, several fixed and movable masks, various diagnostics components including beam viewers, 
x-ray beam position monitors, thermocouples, and vacuum gauges, as well as vacuum valves to separate different vacuum 
sections and to safeguard the vacuum of the storage ring in case of vacuum loss in the beamline. The fixed masks remove 
a vast portion of the undulator radiation power, with the front end typically passing through \(\approx 130\)~W of radiation to the 
optics hutch at the projected 500~mA ring current. The movable mask, based on two L-shaped GLIDCOP slits with tantalum edges 
and located \(\approx 19.5 \) m downstream of the source, is used to define the angular acceptance of photon beam for the 
ForMAX beamline.

\subsection{Primary optics}
ForMAX's primary optics consist of a double crystal monochromator provided by FMB Oxford, a double multilayer monochromator 
by Axilon, dynamically bendable vertical and horizontal focusing mirrors in Kirkpatrick-Baez geometry by IRELEC, a photon 
shutter by Axilon, and four diagnostics modules by FMB Oxford that host a fixed mask limiting the beamline's  acceptance 
angle to \(\leq 24 \times 36~\mu \)rad\(^2 \) (\(x \times y \)) and a high-band-pass diamond filter for heat-load  
management, a white-beam stop, bremsstrahlung collimators, slits, beam viewers, and beam intensity monitors. In the 
following we will briefly discuss the monochromators and mirrors.

\subsubsection{Monochromators}

Depending on the experimental needs, ForMAX can be operated using either a double crystal monochromator (DCM) or a 
double multilayer monochromator (MLM). In line with several other hard x-ray beamlines at MAX 
IV~\cite{ursby20,johansson21,kahnt21}, we have chosen a horizontal deflection geometry for both monochromators to 
maximize their stability. Owing to the relatively high x-ray energy at ForMAX, the energy-dependent polarization factor is \(> 0.75 \) and  
\(> 0.99 \) in the full energy range for the DCM and MLM, respectively. 
In order to facilitate switching between monochromators, both employ the 
same fixed-exit design with 10 mm inboard offset. 

The horizontally deflecting Si(111) DCM is positioned 27 m from the source. We note that the small horizontal offset 
between the crystals allows for a compact and rigid design with excellent stability, as shown 
elsewhere~\cite{kristiansen16}. In ForMAX's case, the 50 mm long upstream crystal is mounted directly on the Bragg goniometer 
(\emph{Ry}) without any other motorized axes, while the 100 mm long downstream crystal has additional motorized adjustments for 
pitch \emph{Ry}, roll \emph{Rz}, and perpendicular motion. Moreover, the monochromator is equipped with motorized 
lateral \emph{x} and vertical \emph{y} translations. Both crystals are side cooled by clamping the crystals to 
liquid-nitrogen-cooled Cu blocks; the high heat load of the upstream crystal requires direct cooling of the Cu block, 
while indirect cooling of the Cu block by braids is sufficient to manage the lower heat load of the downstream crystal. 
  
The horizontally deflecting MLM, in turn, is positioned 25 m from the source. While it is foreseen to be used almost 
exclusively for full-field imaging experiments requiring high temporal resolution, it may also find use in niche, 
photon-hungry scattering experiments. Both multilayer mirrors consists of flat, 180 mm long Si(100) substrates covered with 
separate stripes of 200 layers of W/B\(_4\)C and 250 layers of Ru/B\(_4\)C. Each multilayer stripe has a period of 
\(\approx 2.4\)~nm and \(\approx 1.6\)~nm B\(_4\)C layer thickness, optimized for the energy range of the beamline. 
The bandpass of the MLM, \(\Delta E / E \approx 1\%\) by design, is larger than the width of an individual harmonic peak of 
the undulator. The Bragg rotation \emph{Ry} of the monochromator, fine roll \emph{Rz} of the upstream mirror, and 
fine pitch \emph{Ry} of the downstream mirror are all realized by linear actuators and special flexure arrangements. 
Due to the large angular range of the monochromator, a longitudinal \emph{z} translation of the downstream 
multilayer assembly is needed. Moreover, the motorized motions include the lateral \emph{x} and vertical \emph{y} 
translations of the monochromator as well as the perpendicular translation of the downstream multilayer assembly. Due 
to the significantly smaller Bragg angle compared to the DCM, and hence a larger x-ray beam footprint, it suffices 
to cool both multilayer mirrors indirectly by braids from water-cooled Galinstan baths.

\subsubsection{Mirror system}

The mirror system consists of vertically (VFM) and horizontally (HFM) focusing (and deflecting) mirrors in Kirkpatrick-Baez 
geometry, housed inside a single vacuum chamber. Each mirror has a 650~mm optical length and works at a fixed incidence angle 
of 3 mrad. The mirrors serve two main purposes. First, they provide harmonic rejection. In order to cover the wide 
energy range of the beamline, each mirror from Insync has three separate stripes of Si, Rh, and Pt. Second, each mirror 
can be independently bent to radii between \(\approx 5\) and 100 km, allowing us to focus at the nominal sample 
position or any position downstream thereof, collimate the beam, or essentially operate without focusing. In practice, 
the mirror bending is achieved by applying two controlled bending moments (monitored by strain gauges) at the upstream 
and downstream ends of the mirror in a four-point bending configuration. Each mirror is equipped with a limited amount of stiff, 
motorized axes to maximize stability; lateral \emph{x} and vertical \emph{y} granite translation stages as well as pitch rotation 
(\emph{Rx} for VFM, \emph{Ry} for HFM) employing a high-resolution actuator and flexure parts. Moreover, the HFM 
is also equipped with a similar motorized roll rotation \emph{Rx} by combining a high-resolution actuator and flexure parts.

Mirror slope errors cause striation of the downstream x-ray beam, that is a nuisance when operating the beamline 
in unfocused mode during SR\(\mu\)CT experiments. In order to minimize this effect, each bender is equipped with a set 
of five spring actuators or so-called shape compensators. The residual slope errors for the flat geometry are 
\(\approx 0.11\) and \(\approx 0.13~\mu\)rad for the VFM and HFM, respectively. In the nominal elliptical 
shape for focusing, the mirrors show residual slope errors \(\leq 0.19~\mu\)rad for each stripe.

\subsection{Experimental station}

\begin{figure}
\includegraphics[width=10cm]{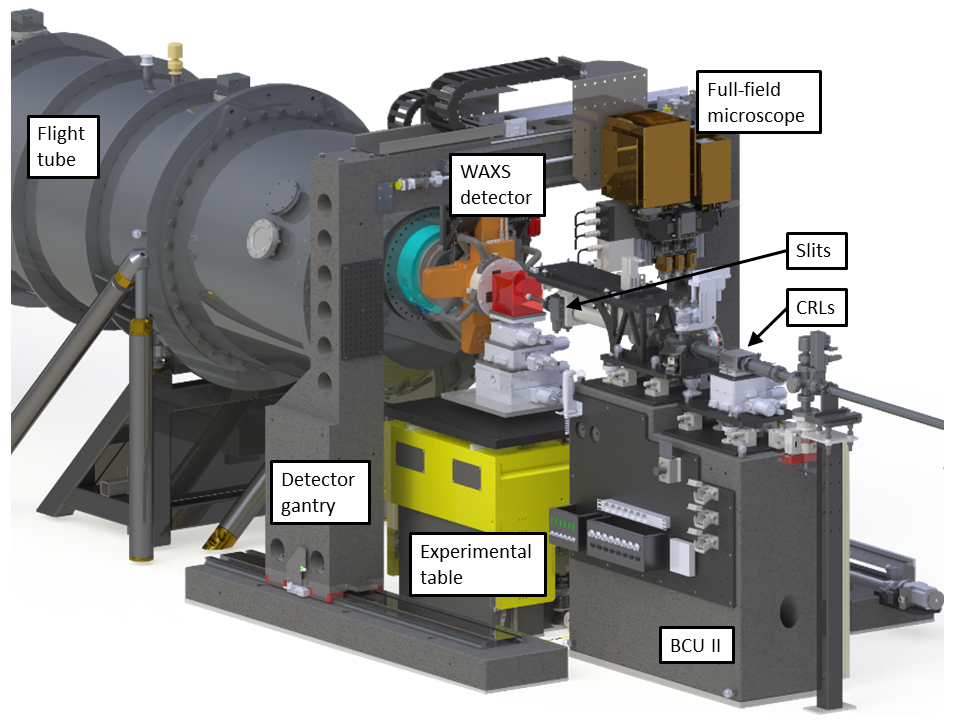}
\caption{Experimental station of the ForMAX beamline. The main components shown include the downstream BCU II, the 
experimental table, the detector gantry, with the WAXS detector and full-field microscope mounted onto it, and the 
evacuated flight tube, hosting the SAXS detector. The CRLs and the monochromatic slits of BCU II are also highlighted for convenience. 
In the SWAXS setup depicted in the figure, the vacuum nose cone, 
onto which the WAXS detector is mounted, is connected to the flight tube using a bellow, while the full-field 
microscope is translated out of the x-ray beam. 
}
\label{FIG:end_station} 
\end{figure}

The major components of the experimental station shown in Fig.~\ref{FIG:end_station} - two beam-conditioning units 
(BCUs), an experimental table, a detector gantry, and a flight tube - have been custom designed at MAX IV. Due 
to the different, and some times mutually competing, technical requirements of SWAXS and SR\(\mu\)CT as outlined above, 
we gave special attention to the integration of these components into a single instrument. Because of the modular 
nature of the experimental station, as described below, we have installed a dedicated programmable logic controller (PLC) system to ensure its 
safe operation. In order to mitigate the effect of parasitic scattering in the small-angle regime, that hampers SAXS studies of weakly scattering 
biobased materials such as low-concentration suspensions of cellulose nanoparticles, all windows in the x-ray beam 
path of the ForMAX beamline are single crystalline. Finally, we have dedicated space between the BCUs to assemble a setup for x-ray multi-projection 
imaging (XMPI)~\cite{Villanueva18,villanueva23}.    

\subsubsection{Beam-conditioning units}
The experimental station hosts two BCUs, positioned approximately 36 and 41 m downstream of the source. The upstream 
BCU (called BCU I) includes a fast shutter, a pneumatic filter unit, and a set of monochromatic slits. In the 
near future, it will also host an x-ray prism lens that allows beam expansion in the \(\approx 5\)~mm range for 
full-field tomographic imaging experiments. The downstream BCU (BCU II) includes a beam viewer, two Si diodes for 
x-ray beam flux monitoring, a set of monochromatic slits, a set of compound refractive lenses optimized to provide 
a microfocus x-ray beam at 16.3 keV for scanning SWAXS experiments, and the possibility of mounting a simple off-axis 
optical microscope for visual monitoring of the sample. In order to minimize the x-ray path in air for 
different setups, the exit vacuum window of BCU II is  motorized along the beam path. Finally, all slits in the experimental 
station are so-called hybrid scatterless slits~\cite{li08}, with single crystal InP wafers mounted on tungsten carbide blades in 
order to suppress parasitic x-ray scattering.

\subsubsection{Experimental table}
The experimental table is located 42 m downstream of the source and is based on a concept developed at the 
ALBA synchrotron~\cite{colldelram10}. The table provides flexibility for sample environment mounting in terms of 
available top surface of \(800\times 800\)~mm\(^2\), load capacity of 200~kg, large lateral and vertical 
translation ranges of 200~mm each, and up to \(\approx 520\)~mm space between the top surface and the x-ray beam.  

The base of the experimental table is a stable and stiff granite block. For vertical \(y\) motion of the table 
(\(\approx 0.3~\mu\)m resolution) we make use of two (upstream and downstream) motorized steel 
plates, that are driven by ball screws with linear guides and actuated by stepper motors. Flexure hinges on the 
steel plates allow fine tuning of the pitch \(Rx\) (20 mrad range,  \(\approx 0.4~\mu\)rad resolution). We have added 
the lateral \(x\) motion (\(\approx 0.3~\mu\)m resolution) on top of the assembly, again driven by a ball screw 
with linear guides.

\subsubsection{Detector gantry}
The granite detector gantry, located by the experimental table, hosts the WAXS detector and the full-field microscope 
for SR\(\mu\)CT. It has five independent motions, \emph{viz.}, 
\begin{itemize}
\item the longitudinal motion of the gantry along the x-ray beam path (\(\approx\)1500 mm range, \(10~\mu\)m resolution), 
\item lateral (\(\approx\)700 mm) and vertical (\(\approx\)20 mm) motions of the WAXS detector (\(10~\mu\)m resolution 
each), and  
\item lateral (\(\approx\)700 mm) and vertical (\(\approx\)30 mm) motions of the full-field microscope (\(1~\mu\)m 
resolution each). 
\end{itemize}
We have verified by measuring the vibrations of the microscope tip with a laser Doppler vibrometer (one minute average, integrated 
4-100 Hz) that the amplitudes are \(<20\)~nm in both lateral and vertical directions (root mean square; RMS).

The above motions permit easy and independent movement of the WAXS detector and the full-field microscope in and out 
of the x-ray beam, thus providing a number of different experimental modes:
\begin{itemize}

\item In the SWAXS setup  (see Fig.~\ref{FIG:end_station}), we center the x-ray beam on the WAXS detector, while the 
SAXS signal (and unscattered 
beam) passes through the central hole of the WAXS detector and impinges on the SAXS detector (and the central beam 
stop). In this setup, we translate the full-field microscope out of the x-ray beam path. In the SAXS setup, in turn, 
we also translate the WAXS detector out of the x-ray beam path and mount an evacuated nose cone onto the flight 
tube to minimize the air path downstream of the sample.     

\item In the SR\(\mu\)CT setup, we translate the WAXS detector out of the path of the x-ray beam. As a safety measure 
we close a gate valve at the entrance of the flight tube, to avoid x-ray exposure of the SAXS detector. 

\item In the combined SAXS and SR\(\mu\)CT setup, we align the SAXS detector with the x-ray beam, and translate the 
full-field microscope vertically in and out of the x-ray beam path for full-field imaging and scattering modes, 
respectively. The vertical translation of the microscope out of the x-ray beam path takes \(\approx 15\)~seconds. A 
combined SWAXS and SR\(\mu\)CT setup is also possible, but the accessible SAXS and WAXS angular ranges are 
limited by space restrictions. 

\end{itemize}

\subsubsection{Flight tube}
In order to minimize (i) absorption of the scattered x-ray beam and (ii) parasitic x-ray scattering from air, we have 
placed the SAXS detector on a motorized detector trolley inside a 9~m long and 1~m diameter evacuated vacuum vessel 
operating at \(\approx 10^{-3}~\)mbar. We have also mounted a motorized central beam stop, made from tungsten  
and equipped with a GaAs diode for monitoring the flux of the transmitted x-ray beam, on the detector trolley. The 
motorized longitudinal motion of the detector trolley permits easy switching of the nominal sample-to-detector distance 
in the range of \(\approx 800 - 7600\)~mm, while the independent motorized lateral and vertical motions allow users to 
freely position the SAXS detector with respect to the direct x-ray beam. We have mechanically decoupled the rail system 
of the detector trolley from the vacuum vessel, thereby isolating the trolley motion from vibrations and 
vacuum-induced deformations of the vessel.

\subsection{Sample manipulation}
\begin{figure}
\includegraphics[width=14cm]{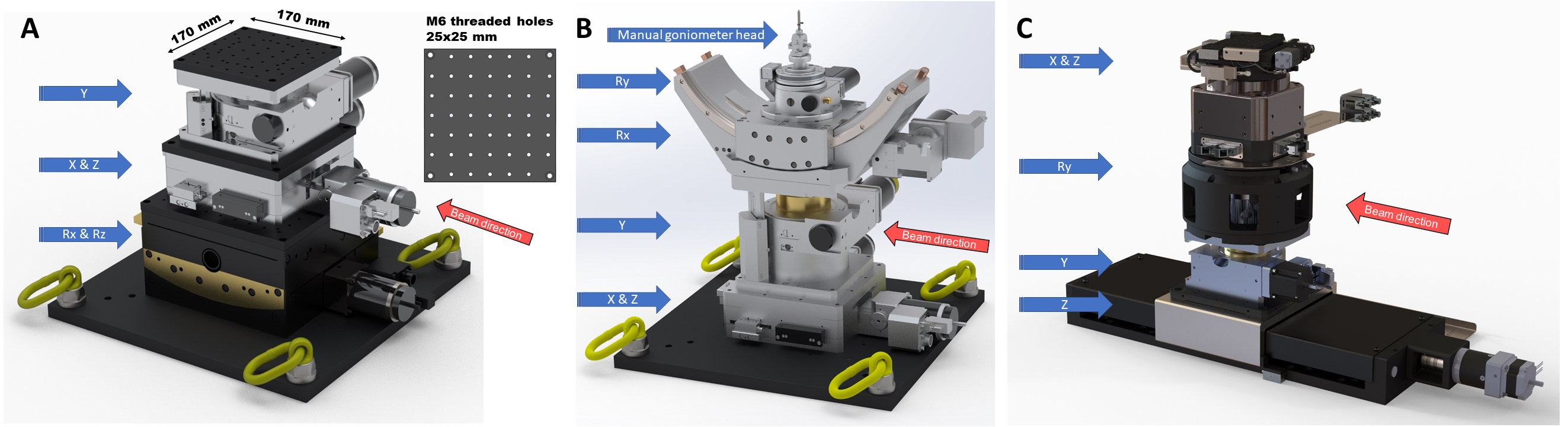}
\caption{Assembly of stages for sample manipulation in (A) SWAXS, (B) scanning SWAXS, and (C) SR\(\mu\)CT experiments. 
}
\label{FIG:sample_stages} 
\end{figure}

ForMAX offers a number of experimental techniques, each with specific requirements with respect to sample manipulation. 
In order to meet different user needs, ForMAX is equipped with three separate stacks of stages for sample 
manipulation:
\begin{itemize}
\item For SWAXS experiments, we provide a high-load (\(\leq 1500~\)N) five-axis assembly from Huber as shown 
in Fig.~\ref{FIG:sample_stages}A. 
It consists, from bottom to top, of motorized pitch \emph{Rx} (\(\pm 13^\circ\)), roll \emph{Rz} (\(\pm 12^\circ\)), 
lateral \emph{x} (\(\pm 25~\)mm), 
longitudinal \emph{z} (\(\pm 25~\)mm), and vertical \emph{y} (\(\pm 20~\)mm) axes. In order to simplify mounting of 
sample holders or environments, we have added an optical breadboard with a \(25 \times 25~\)mm\(^2 \) grid of centered 
ISO metric M6 threaded holes on top of the stages. The nominal distance between the top surface and the center of rotation is 
\(49 \pm 20~\)mm, but this can be increased owing to the modular nature of the assembly of stages.  

\item For scanning SWAXS experiments, we provide another assembly with five degrees of freedom by Huber, see 
Fig.~\ref{FIG:sample_stages}B. The base consists of motorized lateral \emph{x} (\(\pm 25~\)mm), vertical \emph{y} 
(\(\pm 10\) or \(\pm 45~\)mm, depending on resolution and speed requirements), and longitudinal \emph{z} 
(\(\pm 25~\)mm) axes for 2D scanning and adjustment of the sample along the x-ray beam path. On top of these we 
have mounted a yaw \emph{Ry} axis that, combined with the translation stages below, allows  SWAXS tomographic 
imaging. Finally, we have added a large-range, custom pitch axis \emph{Rx} (\(\pm 45^\circ\)) for SWAXS tensor 
tomography experiments. A manual five-axis goniometer head (Huber, model 1002 or 1005) on top of the assembly enable fine 
alignment of the sample.

\item In SR\(\mu\)CT experiments we employ a five-axis assembly from Lab Motion as shown in Fig.~\ref{FIG:sample_stages}C. 
It consists, from bottom to top, of a motorized longitudinal \emph{z} axis (\(\approx 380~\)mm range) for 
propagation-based phase-contrast imaging, a vertical \emph{y} axis (\(\pm 20~\)mm) for helical imaging, an 
air-bearing tomographic yaw axis \emph{Ry}, coupled with a rotary union accomodating a fluid slip ring, and 
horizontal \emph{xz} axes (\(\pm 5~\)mm each) for sample alignment. The electrical slip ring is equipped with 15 
spare wires for integration of sample environments. The maximum rotation speed of the yaw stage is 720 revolutions 
per minute, allowing SR\(\mu\)CT experiments with temporal resolution up to \(\approx 20~\)Hz. The assembly is modular 
and is typically operated without the vertical axis and the rotary union. In order to facilitate mounting of sample 
holders or environments, we have installed an optical breadboard with  a \(12.5 \times 12.5~\)mm\(^2 \) grid of 
centered ISO metric M6 threaded holes on top of the stages. 
\end{itemize}
We further note that we can combine the air-bearing tomographic rotation stage with linear scanning SWAXS stages in 
a modular setup, allowing combined high-resolution SR\(\mu\)CT and 2D/3D scanning SWAXS experiments without the need 
to re-mount the sample upon changing experimental modality.   

\section{Data acquisition and processing}

Data acquisition and processing greatly affects the user experience. In the following, we briefly review how these 
are managed at the ForMAX beamline.    

\subsection{X-ray detection systems ~\label{sec:xraydet}}
 
For SWAXS experiments, ForMAX is equipped with two megapixel, hybrid photon-counting detectors that provide high 
resolution, high dynamic range, and low noise. The SAXS detector is a vacuum-compatible Dectris EIGER2 X 
4M~\cite{donath23}. The WAXS detector, in turn, is a custom X-Spectrum Lambda 3M~\cite{pennicard13}. In 
Table~\ref{TAB:detectors} we provide technical details about both detectors. 

The range of scattering vector modulus \emph{q} covered by the SAXS detector depends on the x-ray energy, 
sample-to-detector 
distance (SDD), positioning of the SAXS detector in the scattering plane, and the size of the central beam stop (at the 
moment 4-5 mm diameter); assuming that the SAXS detector is centered on the direct x-ray beam, the accessible SAXS \emph{q} 
range varies from \(q \approx 0.01 \dots 0.5\)~nm\(^{-1}\) at minimum x-ray energy and maximum SDD to  
\(q \approx 0.25 \dots 10\)~nm\(^{-1}\) at maximum energy and minimum SDD. 

\begin{figure}
\includegraphics[width=8.5cm]{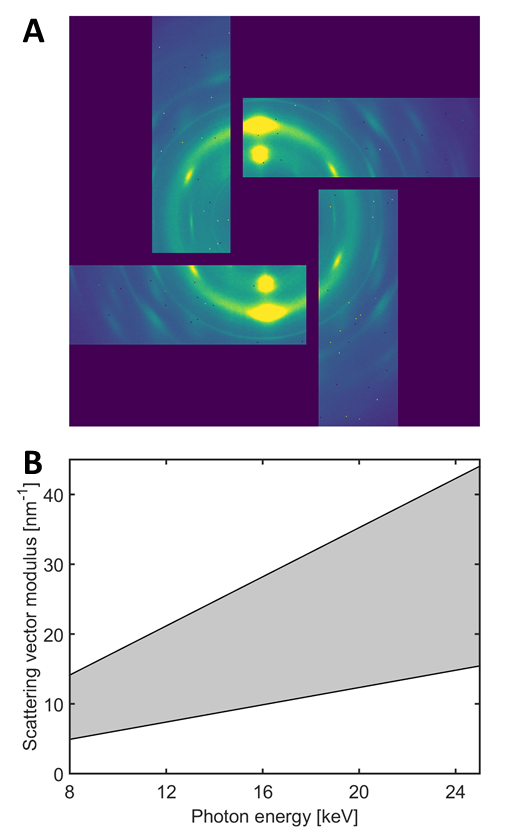}
\caption{WAXS at ForMAX using the custom Lambda 3M 'windmill' detector. Panel~A exemplifies a diffraction pattern from 
a piece of wood, while panel~B shows the nominal accessible range of scattering vector moduli \emph{q} (gray area) 
versus x-ray energy.}
\label{FIG:WAXS} 
\end{figure}

The custom WAXS detector warrants a more detailed discussion. In order to facilitate scanning SWAXS experiments from fibrous 
materials, it has a hole in the center to pass through the SAXS signal (and the direct x-ray beam), while 
simultaneously allowing us to collect WAXS data in all directions of the scattering plane as exemplified in 
Fig.~\ref{FIG:WAXS}A. Moreover, it is mounted onto an evacuated nose cone and connected to the flight tube via a 
bellow, thereby minimizing parasitic air scattering in the SAXS regime. At the nominal SDD of 135 mm, we can collect 
WAXS data at scattering angles \(2\theta = 7 \dots 20^\circ \) in all directions of the scattering plane, yielding 
the energy-dependent range of accessible scattering vector moduli \(q = (4\pi / \lambda) \mathrm{sin}(\theta)\) shown 
in Fig.~\ref{FIG:WAXS}B. We note that there is \(\approx 100\)~mm path of air between the sample and the entrance 
window of the flight tube when using the WAXS detector, adding to the parasitic background scattering in the SAXS regime. Finally, due to the 
thickness of the full-field microscope, SDD\(\geq 235\)~mm in the combined SWAXS and SR\(\mu\)CT experiments, essentially 
halving (i) the energy-dependent minimum and maximum {\em q} of Fig.~\ref{FIG:WAXS}B and (ii) the accessible scattering 
angles \(2\theta\) in the SAXS regime due to shadowing of the flight tube entrance window, hence in practice limiting 
these experiments to x-ray energies \(\geq 20\)~keV. 

For SR\(\mu\)CT experiments, ForMAX is equipped with a high-resolution full-field microtomography detection system 
encompassing two main components $-$ an optical microscope and a sCMOS camera. The transmitted x-ray beam is converted 
by a scintillator into visible light, that in turn is magnified by the optical microscope and recorded by the sCMOS 
camera. The white-beam optical microscope from Optique Peter has motorized triple objective lens and dual camera port 
configurations for simple switching of magnification and sCMOS camera, respectively. We can operate the microscope 
with \(2\times\), \(5\times\), \(7.5\times\), \(10\times\), and \(20\times\) objectives, depending on the required effective 
pixel size and field of view.    

Currently we employ two sCMOS cameras for high-resolution imaging at limited speed, the Hamamatsu ORCA Lightning and 
the Andor Zyla. We are also equipped with a high-speed Photron FASTCAM Nova sCMOS camera to allow 
\(\approx 20\)~Hz SR\(\mu\)CT, i.e., the maximum temporal resolution allowed for by the tomographic rotation stage. 
We summarize the technical details of the sCMOS cameras in Table~\ref{TAB:cameras}.

\begin{figure}
\includegraphics[width=10cm]{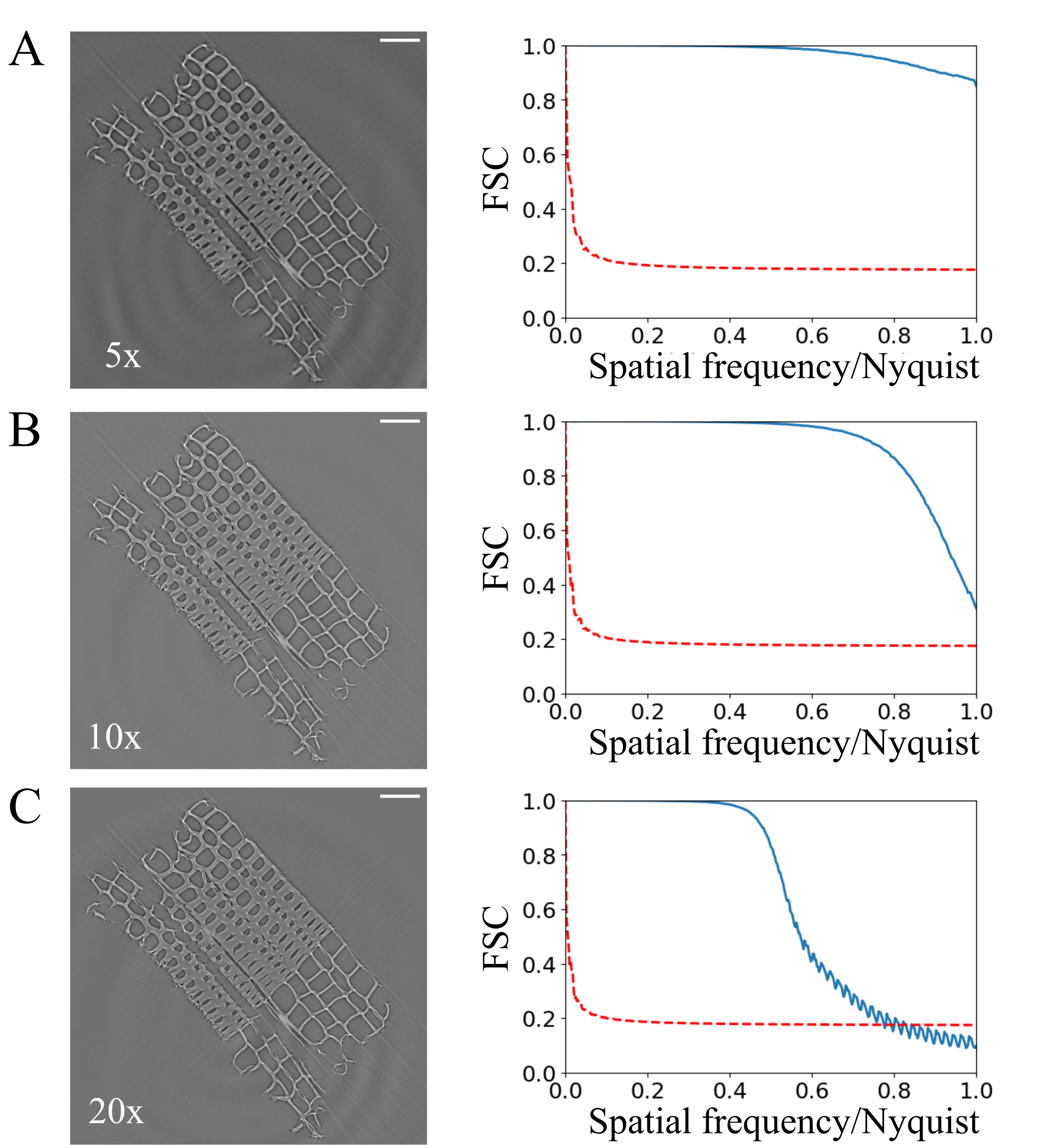}
\caption{SR\(\mu\)CT resolution evaluation at ForMAX. The left column contains the reconstructed slices for (A) \(5\times\), (B) \(10\times\), and 
(C) \(20\times\) magnification (\(100~\mu\)m scale bars). The right column contains the results of the Fourier Shell Correlation 
(FSC) versus spatial frequency (normalized by the Nyquist frequency) for each of the magnifications (blue curve) and the half-bit 
error curve (dashed-red curve).}
\label{FIG:FSC} 
\end{figure}

We have evaluated the performance and resolution of the presented SR\(\mu\)CT system. 
For this evaluation, we used the full-field microscope with \(5\times\), \(10\times\), and \(20\times\) magnification coupled 
to the Andor Zyla camera (physical pixel size 6.5~$\mu$m), resulting in 1.3, 0.65, and 0.325 $\mu$m effective pixel size, respectively. 
The reconstructed slices for the different magnifications of a wood sample are presented in the right column of Fig.~\ref{FIG:FSC}. 
We also evaluated the resolution over the 3D reconstructed volume using the Fourier Shell Correlation (FSC) together with the 
half-bit threshold criterion~\cite{VanHeel87,VanHeel2005} as depicted in the right column of Fig.~\ref{FIG:FSC}. 
We observe that the ForMAX instrument retrieves Nyquist-limited resolution for the \(5\times\) and \(10\times\) magnifications, i.e., 
2.6 and 1.3~$\mu$m resolution. For the \(20\times\) magnification, the retrieved resolution was around 3 pixels, which 
corresponds to 0.975~$\mu$m. Thus, the ForMAX instrument is ideal to characterize objects in 3D with 
micrometer resolution.

\subsection{Control system}

ForMAX's control system is based on Tango~\cite{chaize99}, an open-source control system that is in use at several 
European synchrotron facilities. On top of Tango we employ Sardana~\cite{coutinho11}, a software environment for, 
e.g., controlling motors, acquiring signals, and running macros. We have optimized the scan routines for the specific 
needs of ForMAX, such as reducing the overhead per line in continuous \emph{xy} mesh scans to \(< 1\)~second for 
scanning SWAXS applications. From a hardware point of view, the majority of our motorized axes are based on stepper 
motors controlled by IcePAP~\cite{janvier13} and we make use of PandABoxes for synchronization of the 
experiments~\cite{zhang17}.

\subsection{Data pipelines}

All detectors are integrated into the beamline control system via dedicated detector servers, utilizing 
detector-specific software development kits (SDKs) running on detector control units (DCUs) for detector control 
and data readout. Low-level image processing such as flat-field correction is either applied by default (for hybrid 
pixel detectors; SWAXS) or in the image reconstruction (for sCMOS cameras; SR\(\mu\)CT), while low level acquisition 
parameters such as acquisition time, number of frames, and photon-counting threshold energy are accessible to the 
user. Finally, the data are streamed via high-speed Ethernet connections to MAX IV's central data 
storage and saved together with metadata in the hierarchical data format 5 (HDF5). In the case of the high-speed 
Photron sCMOS camera, the data will temporarily be saved on the local 
data storage of the camera, before transfer to the central data storage. The data are stored for at least seven years. 

\begin{figure}
\includegraphics[width=14cm]{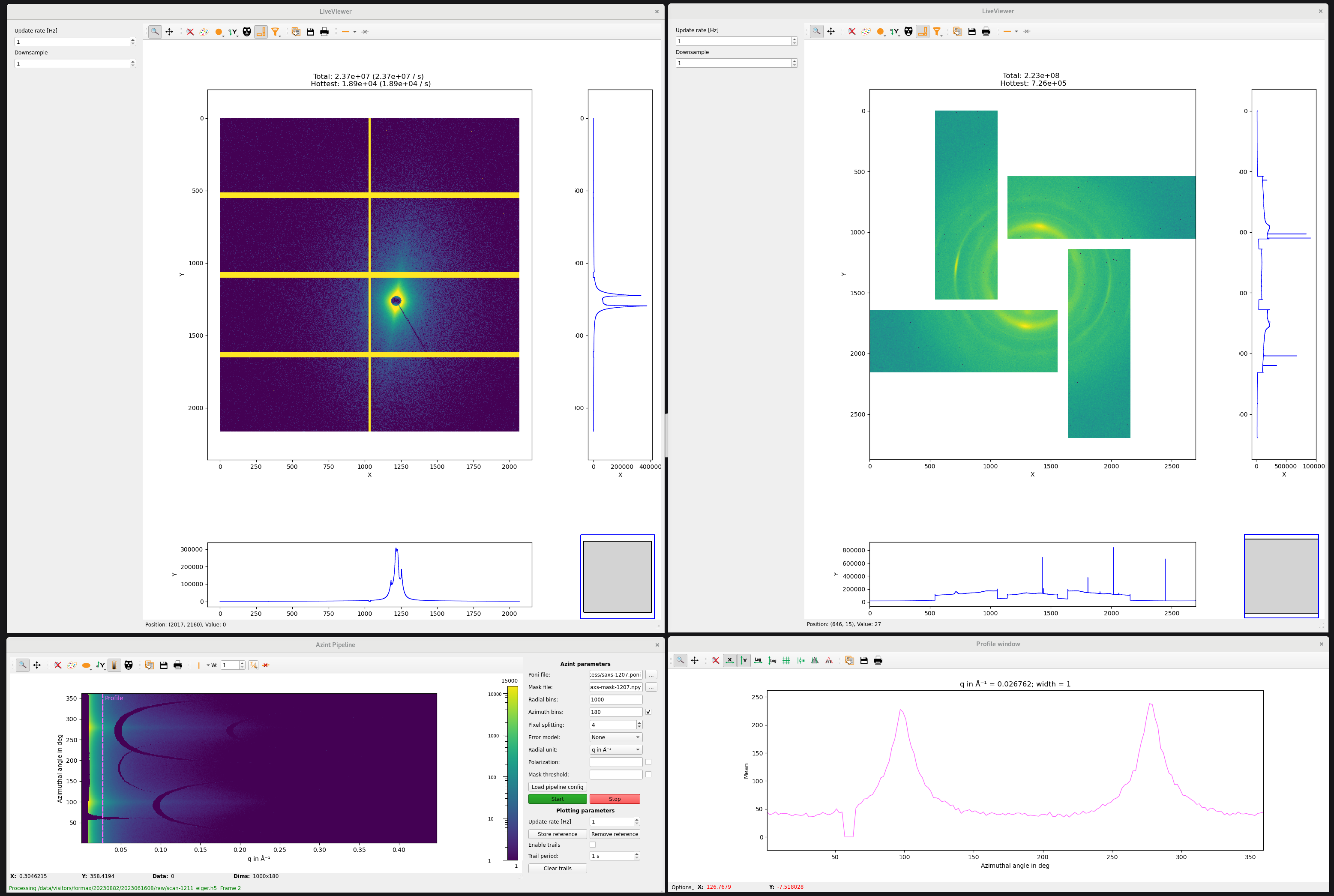}
\caption{Snapshot from the beamline control computer, exemplifying the live plotting of SWAXS data. The top panel 
shows as measured SAXS (left) and WAXS (right) data collected from a wood sample. The bottom left panel illustrates 
reduced 2D \(I(q,\varphi)\) data in the SAXS regime. The line profile of the reduced 2D data, shown in the bottom 
right panel, corresponds to an annular integral of the as measured SAXS data and is convenient for monitoring 
anisotropy in the scattering data. Similar live plotting of reduced WAXS data is available at the beamline. 
}
\label{FIG:azint} 
\end{figure}

In parallel with data streaming and storage of the as measured scattering data, our SWAXS data pipeline reduces the 
data to a more user-friendly format. The data reduction is carried out using the python implementation of 
\emph{MatFRAIA} \cite{Jensen22}, based on a matrix-multiplication algorithm for radial and 
azimuthal integration, and is faster than the maximum frame rate of the EIGER detector. We reduce the SWAXS data into both 
1D \(I(q)\) and 2D \(I(q,\varphi)\) formats, where \(I\) denotes the scattering intensity and \(\varphi\) the azimuthal 
angle. We emphasize that the fast data reduction into so-called 'cake plots', 
\(I(q,\varphi)\), is particularly convenient for monitoring anisotropic scattering from fibrous materials in 
(scanning) SWAXS experiments. For calibration and masking of the detectors we utilize \emph{PyFAI} \cite{kieffer18}, 
which is well known in the user community. Finally, in order to facilitate monitoring of the experiment, 
we plot either the radial integral \(I(q)\) or the 'cake plot' \(I(q,\varphi)\) in both SAXS and WAXS regimes in live mode. 
In Fig.~\ref{FIG:azint} we present a snapshot from the beamline control computer, exemplifying the live plotting of 
reduced SWAXS data. 

In SR\(\mu\)CT experiments, we take a different approach for the data pipeline. In line with community convention, 
we collect projections as well as flat- and dark-field images using dedicated scan routines and save all data 
in common HDF5 files. In order to further improve user friendliness, we are currently in the process of implementing 
live tomographic reconstructions for SR\(\mu\)CT experiments.

\subsection{Data analysis and image reconstructions}

ForMAX allows a wide range of SWAXS, scanning SWAXS, and SR\(\mu\)CT experiments, each with their unique requirements 
with respect to on-line data analysis. In order to support all these different experiments, we provide up-to-date 
\emph{Jupyter notebook} templates for our users. For SR\(\mu\)CT experiments, the script for tomographic reconstruction includes the 
option to perform phase retrieval for single-distance propagation-based phase contrast tomography, in addition to standard 
absorption contrast tomography reconstruction. We plan to continuously implement further developments, including the 
aforementioned live tomographic reconstructions for SR\(\mu\)CT experiments.   

SAXS tensor tomography (SASTT), that combines concepts of scanning SAXS with SR\(\mu\)CT to retrieve 
not solely scattering intensity measures but the full 3D reciprocal space map within each voxel of the tomogram, is 
a special case due to the high demands on image reconstruction. Data acquisition must be matched with sufficient 
computational resources to allow reconstructions of the 3D reciprocal space map, ideally already during the experiment, 
to evaluate the quality of the measurements. At ForMAX, we have implemented \emph{Jupyter notebook} templates for projection 
alignment and apply the software package \emph{Mumott} (mumott.org) to perform SASTT 
reconstructions. In the future, we plan to continuously update these notebooks to remain up to date and match further 
developments and improvements of the reconstruction algorithm. Details about \emph{Mumott} can be found in a recent 
publication by \cite{nielsen23}.

\section{Benchmarking}

In this section we report on initial benchmarking of the main x-ray beam properties at ForMAX.  

\subsection{Beam size} 

The dynamically bendable mirrors provide means to vary the lateral and vertical  beam size at the sample position in 
a large range. In the unfocused mode, we obtain an FWHM beam size of \(\approx 0.8 \times 1.3~\)mm\(^2 \) 
(\(x \times y\)) using the DCM, while the larger bandpass of the MLM yields a beam size of 
\(\approx 1.3 \times 1.5~\)mm\(^2 \). In the other extreme, we can focus the 
beam down to \(\approx 55\dots60 \times 10\dots15~\mu \)m\(^2 \) at the sample position using either monochromator. 
In this case, the lateral beam size is limited by source size and imaging geometry, while the vertical beam size 
is limited by slope errors of the mirrors. The dependence of the beam size on the ID harmonic is negligible. 

In order to further decrease the beam size at the sample position, we have installed compound refractive lenses 
(CRLs) \(\approx 1.5\)~m upstream of the nominal sample position. At the moment we employ a stack of 16 
radiation-resistant SU-8 polymeric lenses from Microworks, optimized for 16.3 keV x-rays and yielding a FWHM beam size 
of \(\approx 10 \times 2~\mu\)m\(^2 \) at the sample position. This is similar to the beam size typically available 
at third-generation SWAXS beamlines with microfocus capability~\cite{buffet12,smith21}. 

The natural beam size at ForMAX limits SR\(\mu\)CT experiments on large samples. This limitation can be partly overcome 
by stitching images, but at the expense of temporal resolution. We will soon also install an optional, overfocusing 
SU-8 x-ray prism lens from Microworks \(\approx 5.4\)~m upstream of the nominal sample position, yielding an 
energy-dependent x-ray beam size of \(\approx 5 \times 5~\)mm\(^2 \) or larger at the sample position in 
combination with the DCM.

\subsection{Flux} 

\begin{figure}
\includegraphics[width=8.5cm]{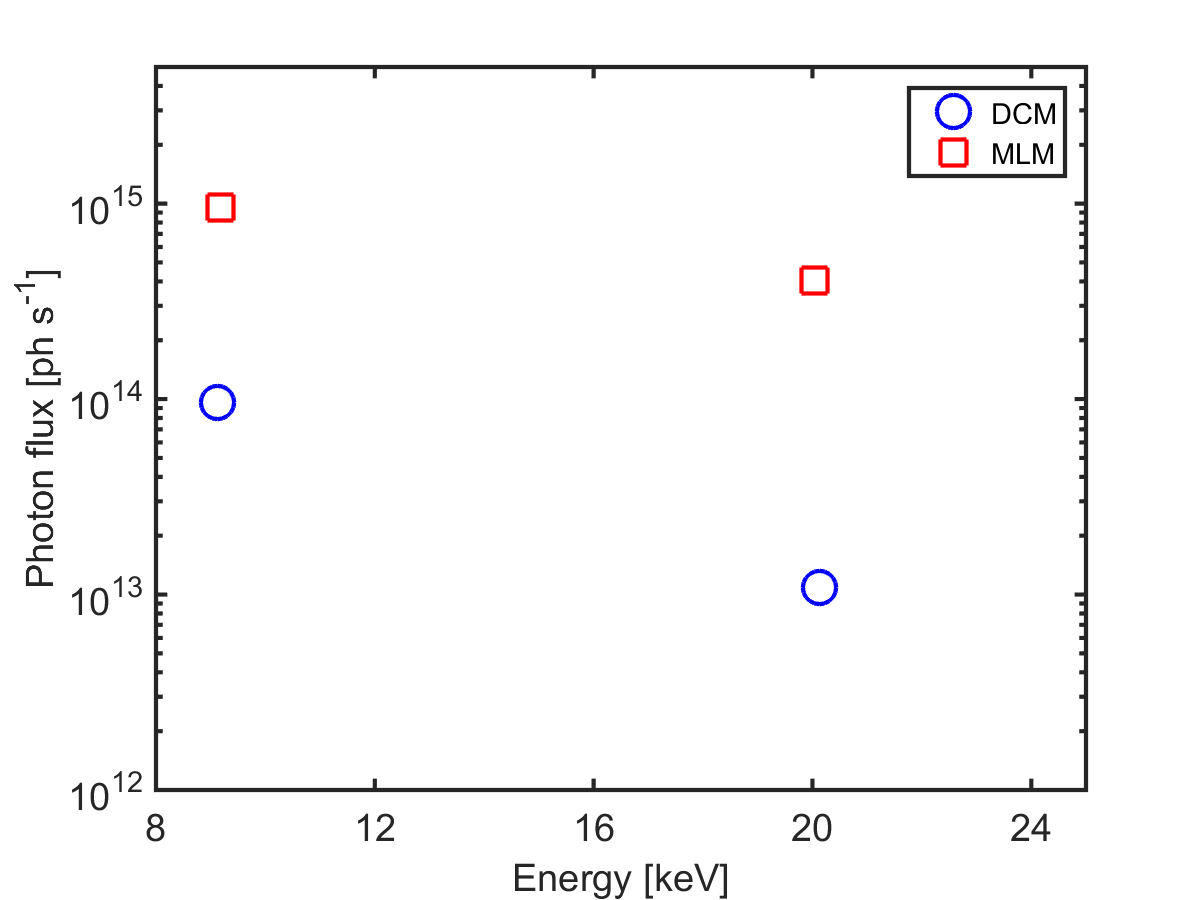}
\caption{Measured x-ray photon flux at the sample position versus x-ray energy, shown for both monochromators and selected x-ray 
energies. The data were obtained using the minimum undulator gap and the maximum, \(24 \times 36~\mu \)rad\(^2 \) (\(x \times y \)) 
acceptance angle, i.e., the typical configuration for photon-hungry SR\(\mu\)CT and scanning SWAXS experiments.}    
\label{FIG:flux} 
\end{figure}

The imaging techniques available at ForMAX rely on a large incident photon flux. In Fig.~\ref{FIG:flux} we present 
the measured x-ray photon flux at the sample position for both monochromators. We collected the data with the minimum 
undulator gap (4.5 mm) and maximum acceptance angle (\(24 \times 36~\mu \)rad\(^2 \)), as typically employed for photon-hungry 
SR\(\mu\)CT and scanning SWAXS experiments. We measured the flux of a strongly attenuated x-ray beam at \(\approx 9\) and 
20~keV using the photon-counting EIGER detector, an approach that yielded reliable results owing to the efficient harmonic rejection 
using the Si and Rh stripes of the mirrors at these energies. The measured fluxes agree within a factor of three with results based on 
ray-tracing simulations with \emph{XRT}~\cite{klementiev14}, assuming ideal undulator and optics. 

Let us briefly discuss the available x-ray flux at ForMAX compared to competitive beamlines at third-generation synchrotron sources. 
In terms of SWAXS, the x-ray flux at the latter for typical experimental conditions in the enrgey range of ForMAX is generally 
\(\approx 10^{13}\)~ph~s\(^{-1}\) or smaller, see e.g. \cite{smith21}. The smaller x-ray beam divergence at ForMAX allows 
for a photon flux (using the DCM) of up to an order of magnitude larger than these values, greatly facilitating scanning SWAXS experiments. 
Moreover, the MLM is available for niche experiments requiring an even higher flux. 
In terms of SR\(\mu\)CT, in turn, the flux at ForMAX (using the MLM) is comparable to one available at third-generation 
sources \cite{stampanoni06,rau11,vaughan20}, albeit in an up to two orders of magnitude smaller beam cross section. We note that while the 
small beam size at ForMAX limits the capability of full-field imaging of large samples, as alluded to above, the very high photon density 
instead carries the potential for ultrafast imaging.  

\subsection{Coherence estimation}
In this section, we present an initial quantification of the coherent properties at the ForMAX beamline. Specifically, we 
evaluate the effects of coherence in the formation of holographic fringes in an in-line holography experiment. 
We envision performing an exhaustive analysis of the coherent properties of ForMAX~\cite{Goodman1985,Vartanyants2010} for 
different energies and imaging configurations, but this is out of the scope of the present paper.

\begin{figure}
\includegraphics[width=8.5cm]{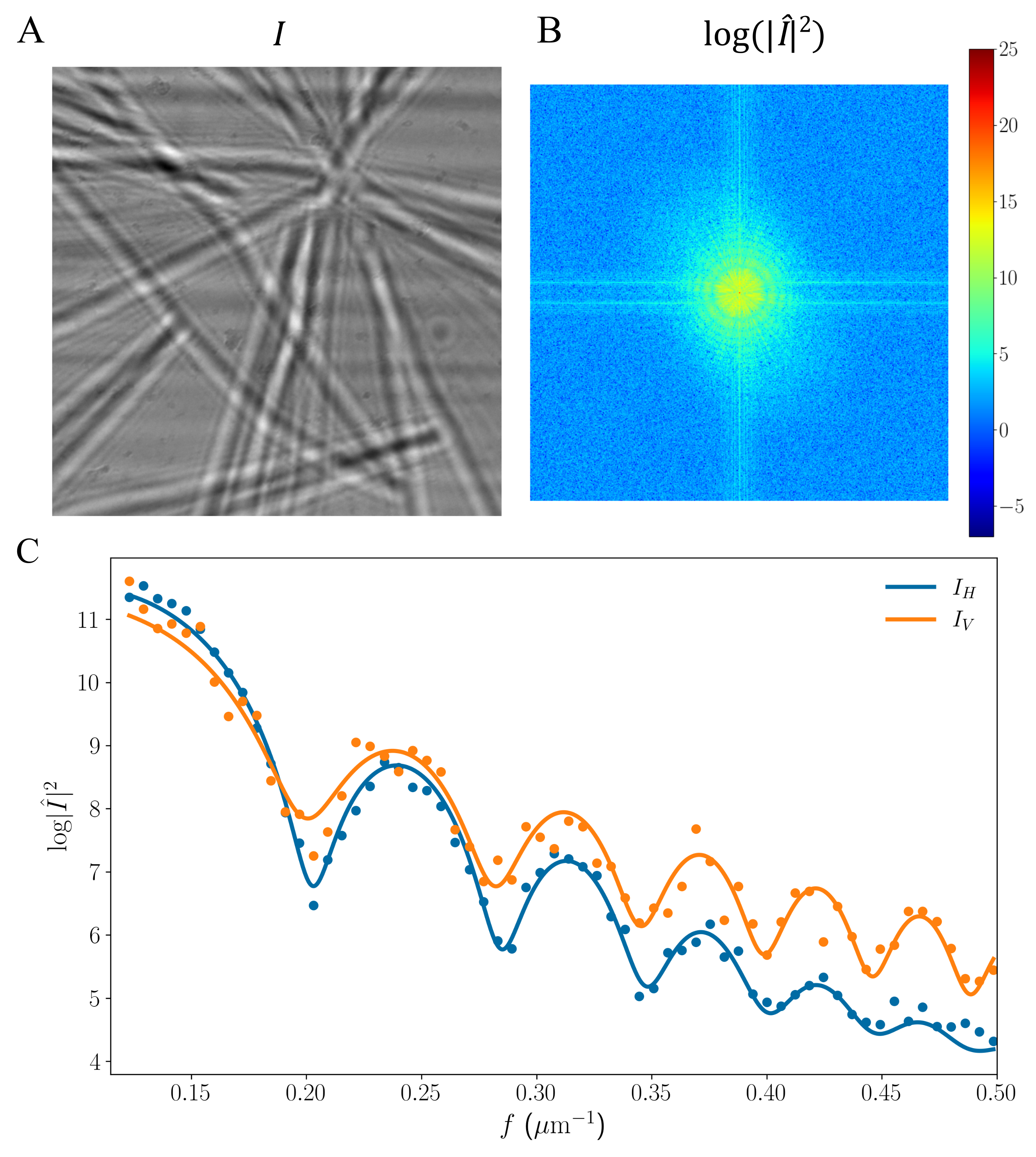}
\caption{Coherence evaluation via contrast-transfer-function analysis (CTF). (A) Hologram recorded from a broken 
$\mathrm{Si_3N_4}$ membrane. (B) Power spectrum of the hologram in logarithmic scale ($\log|\hat{I}|^2$). 
(C) Lateral (blue symbols) and vertical (orange symbols) components of the power spectrum. The solid lines depict fits to the data.} 
\label{FIG:CTFanalysis} 
\end{figure}

We performed in-line holography at 9.1~keV, imaging a broken $\mathrm{Si_3N_4}$ membrane that exhibited  
several sharp edges with random orientations~\cite{dierks20}. Fig.~\ref{FIG:CTFanalysis}A depicts the hologram ($I$) 
recorded 19~cm downstream of the sample, using the SR\(\mu\)CT detection system with an effective pixel size of 0.325~$\mu$m 
and a response function (also known as the Point Spread Function; PSF) with a FWHM corresponding to 3 pixels 
($\sigma_{\mathrm{PSF}}$=0.41~$\mu$m), as estimated in section~\ref{sec:xraydet} 
for the 20x magnification objective. The Fourier transform of the recorded hologram ($\hat{I}$) can be written as~\cite{Zabler2005}
\begin{equation}
    \hat{I}=\left[\delta(f)+2\hat{\phi}\sin(\pi\lambda z f^2) \right]\hat{R}(f)\gamma_C(f), 
    \label{eq:CTF}
\end{equation}
where $f$ is the frequency, $\phi$ the wave's phase after the object, $\lambda$ the wavelength, $z$ the 
propagation distance between the sample and the detector, $R$ the detector's point-spread function, and $\gamma_C$ 
the degree of coherence. The sinusoidal term in Eq.~(\ref{eq:CTF}) is also known as the contrast-transfer 
function~\cite{Guigay1977} and the visibility of its oscillations as a function of the frequency is limited by the 
coherence and the PSF. The power spectrum ($|\hat{I}|^2$) of the recorded hologram in logarithmic scale is 
depicted in Fig.~\ref{FIG:CTFanalysis}B. We clearly observe an asymmetry between the visibility of the CTF 
oscillations in the vertical and lateral directions due to coherent effects.

For an initial quantification of the coherence effects, we performed a fit of Eq.~(\ref{eq:CTF}) to the power spectrum, 
describing the PSF and the degree of coherence by a Gaussian function with the standard deviation
\begin{equation}
    \sigma_{\mathrm{TOT}=\sqrt{\sigma^2_{\mathrm{PSF}}+\sigma^2_C}},
\end{equation}
where $\sigma_C$ is due to the degree of coherence. Because of the difference in phase space of the source in the principal directions, 
we fitted the data independently in vertical and lateral directions as presented in Fig.~\ref{FIG:CTFanalysis}C. 
On the one hand, the vertical $\sigma_{\mathrm{TOT}}\approx 0.40~\mu$m is comparable to $\sigma_{\mathrm{PSF}}$, implying that 
the blurring of the fringes in the vertical direction is dominated by the PSF. This observation is compatible with the small vertical electron 
source size, and hence large vertical coherence length, of Table~\ref{TAB:MAXIV}.  
On the other hand, the lateral $\sigma_{\mathrm{TOT}}\approx 0.71~\mu$m corresponds to $\sigma_C\approx 0.58~\mu$m, 
suggesting that the lateral visibility is mainly limited by the degree of coherence. This finding is in line with  
the larger lateral electron source size, and hence smaller lateral coherence length, of Table~\ref{TAB:MAXIV}.

\section{Probing hierarchical materials}

The objective of the ForMAX beamline is to provide multiscale and multimodal structural characterization of materials 
from nm to mm length scales. In the following, we demonstrate this capability with a few examples. 

\subsection{Combined scanning SWAXS and SR\(\mu\)CT}

\begin{figure}
\includegraphics[width=14cm]{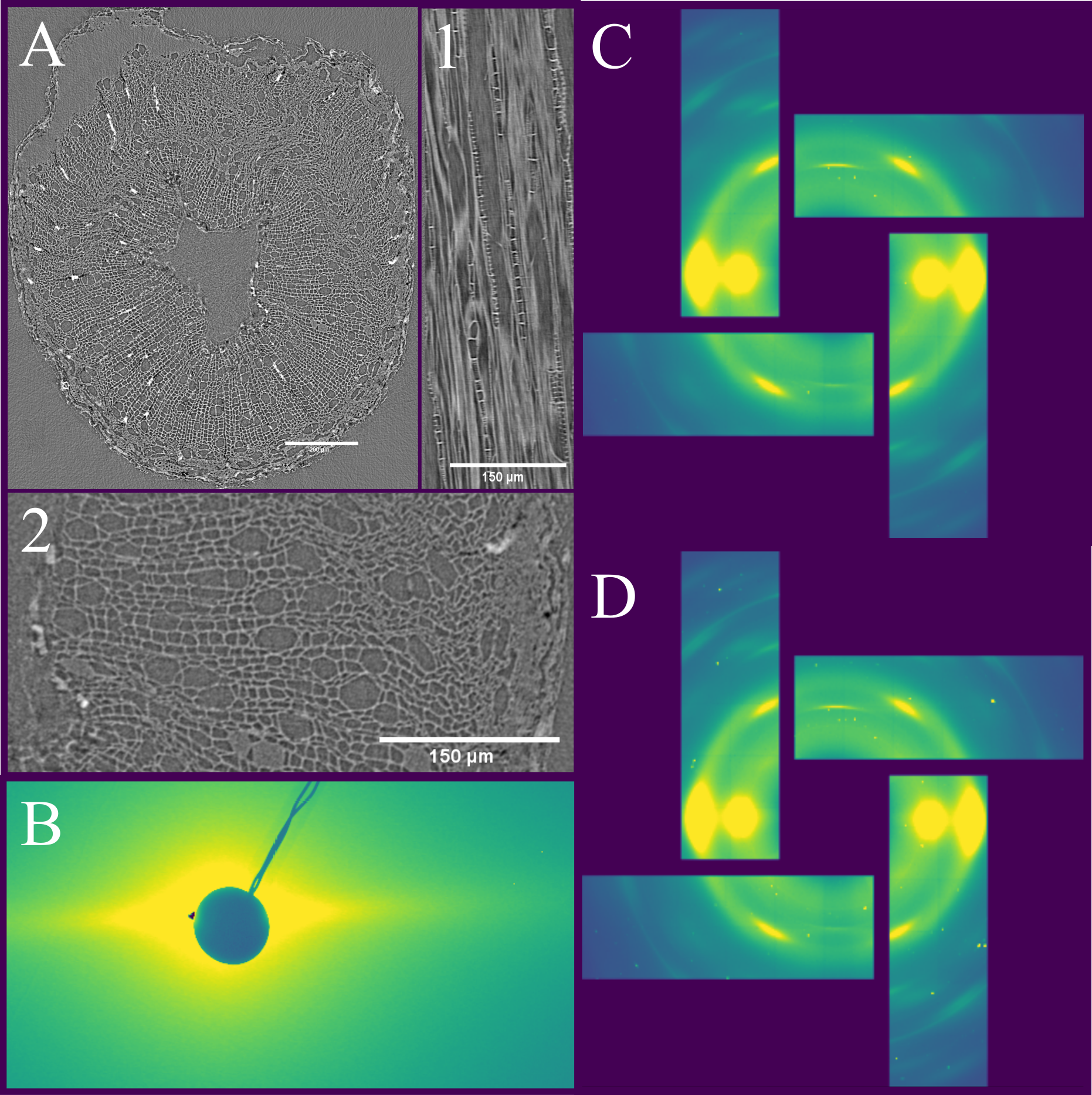}
\caption{Zooming into hierarchical materials at ForMAX. Panel A shows a 2D slice from the reconstructed 3D volume of an aspen sapling 
with a magnified view into cellular structure in both tangential (A1) and radial directions (A2) as obtained by SR\(\mu\)CT. 
Such data provide microscopic structural characterization and allows users to identify regions of interest for nanoscale mapping.
Panels B and C-D present local SAXS and WAXS data, respectively, acquired using an x-ray beam 
focused to \(\approx 25 \times 25~\mu \)m\(^2 \) at the sample position. Panels C and D illustrate spatially 
resolved WAXS mapping of crystallites, measured at different positions within the sample. 
}    
\label{FIG:comb_SWAXS_CT} 
\end{figure}

A key feature of ForMAX is the possibility of zooming into hierarchical materials, as illustrated in 
Fig.~\ref{FIG:comb_SWAXS_CT}. In a first instance, we acquire a high-resolution 3D image by SR\(\mu\)CT, yielding 
microscopic structural characterization of the sample. This is exemplified in Fig.~\ref{FIG:comb_SWAXS_CT}A for a 
sample of aspen sapling mounted in tangential geometry. Moreover, the tomogram allows us to identify regions 
of interest (RoI) for scanning SWAXS mapping of nanoscale structures. In the second instance, we focus the x-ray beam 
onto the sample position and collect spatially resolved SWAXS data on either the RoIs or the full sample, as illustrated 
in Figs.~\ref{FIG:comb_SWAXS_CT}B and C-D for SAXS and WAXS, respectively. 
Here, the anisotropy in the SAXS data is due to scattering from cellulose fibrils, whereas the main WAXS signal is due 
to diffraction from their crystalline parts.  
We note that whereas the SAXS and WAXS data 
of Figs.~\ref{FIG:comb_SWAXS_CT} provide access to structural properties such as microfibril size and orientation, the 
WAXS data of Figs.~\ref{FIG:comb_SWAXS_CT}C and D also allow the mapping of other crystalline compounds within the 
sample, in this case calcium oxalate crystals. For bio-based materials, different crystalline agents are often 
present in the samples, and with the combination of spatial SR\(\mu\)CT and SWAXS data, the regions of interest within 
the sample can be reconstructed using various scattering contrasts. 
 
The feature of zooming into hierarchical materials is still under development. Potential means for improving user 
friendliness include, e.g., a graphical user interface for selecting RoIs from the 3D SR\(\mu\)CT data. Nevertheless, 
ForMAX provides already in its present state unique means for multiscale and multimodal structural characterization 
of soft and/or bio-based materials in the nm to mm range.  
 
\subsection{SAXS tensor tomography}

As noted in the introduction, scanning SWAXS imaging provides structural characterization across seven orders of 
magnitude in length scales in a single experiment. SAXS tensor tomography (SASTT) is particularly useful for 
hierarchical materials, since the statistically averaged local orientation of fibrils, fibers, or filaments accessible 
in these experiments is directly linked to the mechanical properties of the sample. In the scope of commissioning 
the beamline, we acquired a SASTT dataset from carbon fiber bundles that were carefully arranged in the shape of a 
small knot. A similar test sample has already been used in the initial SASTT commissioning experiments at the cSAXS 
beamline, Swiss Light Source \cite{liebi15}. The purpose of such a measurement is to ensure the proper mapping of 
3D reciprocal space (scattering directions) and real space directions, \(xy\) scanning at different rotation and 
tilt orientations, into the reconstruction algorithm. We successfully reconstructed the first dataset already during 
the beamtime, due to the readily available computing resources at the MAX~IV high performance cluster (HPC). 

\begin{figure}
\includegraphics[width=8.5cm]{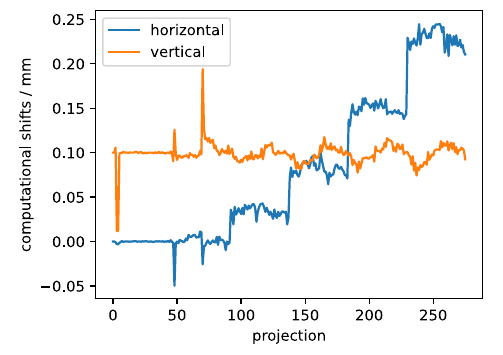}
\caption{Horizontal and vertical shifts as computed by the alignment procedure and applied to all projections before 
used as an input for the SASTT reconstructions via \emph{Mumott}. Shifts are computed with the \emph{phase\_cross\_correlation} 
function from the \emph{skimage.registration} package in python with a Filtered-Back-Projection (FBP) tomogram from 
the measurement at \(0^\circ \) tilt as a reference. 
}
\label{FIG:SASTT_stab} 
\end{figure}

The input for the reconstruction consists of a dataset with 276 two-dimensional projections with 
\(55 \times 76\)~pixels (\(x \times y\)) at a pixel size of \(25~\mu\)m, computed from a total of 
1.15M detector frames. Each pixel of every projection consists of detector data which was reintegrated into 
32 azimuthal bins in the range of \(q = 0.3 \dots 0.5 \)~nm\(^{-1}\) and further symmetrically averaged to remove 
detector gaps. The remaining 16 azimuthal bins are used as input for the SASTT reconstruction with \emph{Mumott} 
version 1.2 (https://zenodo.org/records/8404162). 
Another important step in the workflow is projection alignment. We used a computational procedure to align all 
projections for different orientations of the sample (\(Rx = 0 \dots 180^\circ\) for \(Ry = 0^\circ\),  
\(Rx =0 \dots 360^\circ\) for \(Ry > 0^\circ\)), that first generates a tomogram for \(Ry = 0^\circ\) and next 
back projects the projections of all other tilts and uses \emph{phase\_cross\_correlation} from the 
\emph{skimage.registration} python package for image registration and computation of the required shifts. 
We used the integrated dark field signal as input for the alignment procedure, due to the weak absorption signal from 
the carbon fibers. The same procedure was further used to mask out the sample holder and frame from some of the 
projections. The computed vertical and horizontal shifts are in total \(\leq 300~\mu\)m (see Fig.~\ref{FIG:SASTT_stab}), 
showing that the experimental setup is very stable. 

We reconstruct the 3D reciprocal space in each voxel using band-limited Friedel symmetric spherical functions expressed 
in spherical harmonics up to a maximum order of 6, which results in 28 coefficients for each voxel that are used to 
reconstruct the 3D reciprocal space. The orientation of the main structure is determined from the eigenvector 
associated with the smallest eigenvalue of the rank-2 tensor. We have checked the robustness of the reconstruction by visual 
comparison of 2D orientation, anisotropy, and degree of orientation between the measurements and simulated projections 
of the reconstructed data. Finally, we calculate the degree of orientation as the ratio between the mean (isotropic 
component) and standard deviation (RMS of anisotropic component).

\begin{figure}
\includegraphics[width=14cm]{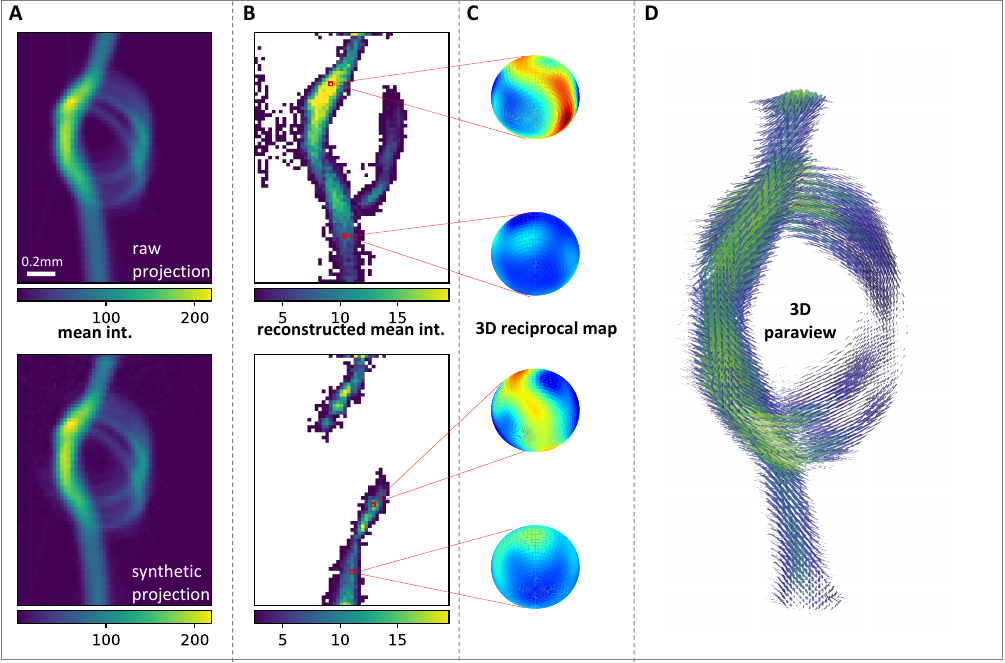}
\caption{Summary of SASTT analysis of a carbon fiber knot. Part A compares the mean intensity of a measured 
projection with the corresponding synthetic projection computed from the reconstruction. In B, two central cuts through 
the tomogram of the mean intensity (\emph{zx} and \emph{zy} slice) visualize the content of fibers throughout the 
tomogram. Four selected voxels are highlighted in red, for which the reciprocal space map is shown in C (interpolated 
with a five degree resolution and projected on a sphere). D presents a \emph{ParaView} rendering of the knot.
}    
\label{FIG:SASTT} 
\end{figure}

We display the results of the reconstruction in Fig.~\ref{FIG:SASTT}. In Fig.~\ref{FIG:SASTT}A, we directly compare 
the input data of the mean intensity with a synthetic projection computed from results of the reconstruction. Since 
there is essentially no difference between measured and synthetic projections, which is the goal of the reconstruction, 
we move on to inspect the tomogram in more detail. Fig.~\ref{FIG:SASTT}B displays two central cuts, a \emph{zx} 
and a \emph{zy} slice through the tomographic reconstructed mean intensity, which gives direct insights into the 
arrangement of the fibers within the knot. The top left of the image exhibits a region of higher intensity where the 
fiber bundles from top/bottom overlap, while the opposing side of the image shows two open loops of less densely 
packed material. Besides the mean intensity, SASTT reconstructions also offer the unique possibility to assess the 
3D reciprocal space map within each voxel, as shown in Fig.~\ref{FIG:SASTT}C for selected voxels of the tomogram 
(scaled with the same colormap for better comparison). The high intensity region clearly shows a ring-like reciprocal 
space map, which is expected for fiber-like structures. Finally, in Fig.~\ref{FIG:SASTT}D we visualize the combined 
information of the carbon fiber knot using the visualization software \emph{ParaView} (www.paraview.org). 
Cylinder glyphs with fixed aspect ratio point in the direction of the carbon fibers. We use the mean intensity, a measure 
of the material's density, to scale as well as color-code the glyphs. Note that we have masked the output with a 3D array 
taken from the mean intensity to exclude low scattering regions and mask out data from air/background voxels. 

\subsection{Advanced rheological and mechanical testing}

\emph{In situ} rheological or mechanical testing is a common approach to address, for example, flow-induced assembly 
of nanoparticles into advanced materials or the relationship between structural and mechanical properties in fibrous 
materials. We foresee that such studies will be popular among our user community. However, while combined rheological 
and small-angle scattering experiments are rather mature~\cite{eberle12}, a deep understanding of flow-induced assembly 
of nanoparticle suspensions into novel hierarchical materials requires simultaneous rheological and multiscale structural 
characterization~\cite{kadar21,kadar23}. This is particularly important for the assembly and development of new materials 
from biomass, for which the importance of flow cannot be overstated. Likewise, while \emph{in situ} uniaxial tensile or 
compressional load is commonly excerted during SWAXS or SR\(\mu\)CT experiments, materials engineering applications may 
require more complex load  geometries and loading profiles or extensive load cycling in well controlled temperature and relative 
humidity. Again, this is of great importance for biobased materials, that are viscoelastic even in their solid state.

\begin{figure}
\includegraphics[width=14cm]{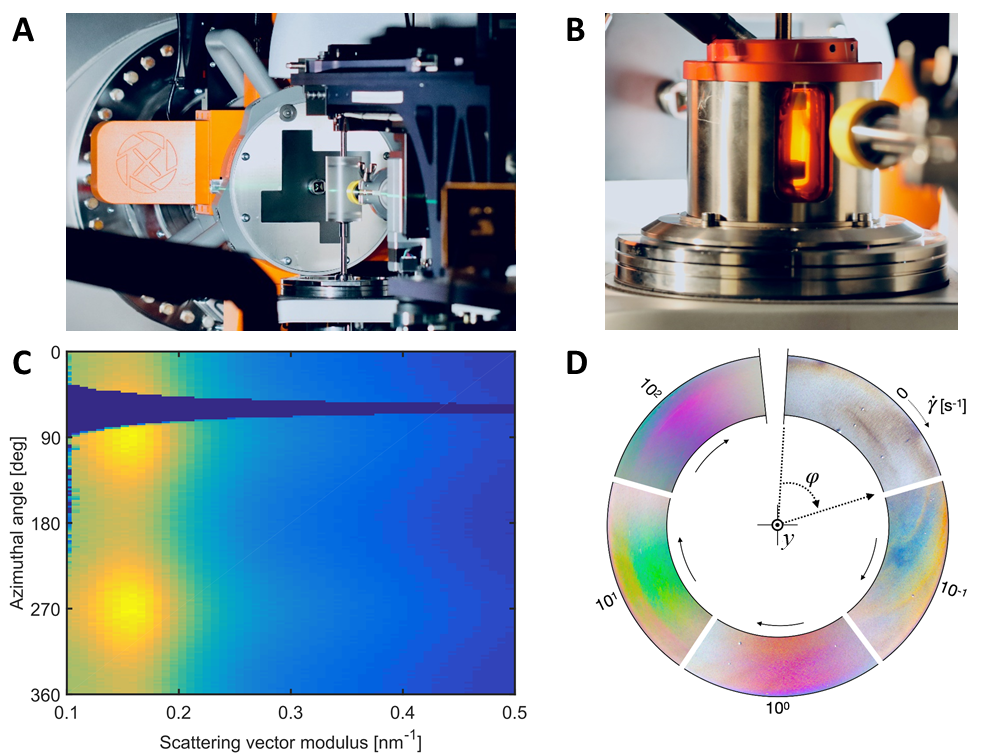}
\caption{Examples of \emph{in situ} rheological and mechanical testing at ForMAX. 
Panels A and B show the Rheo-SWAXS and DMA-SWAXS setups at ForMAX, respectively, based on the Anton Paar MCR702 
rheometer available at the beamline. The SAXS data of panel C collected from a suspension of cellulose nanocrystals 
(CNC), presented as a function of scattering vector modulus \emph{q} and azimuthal angle \(\varphi\), show 
nanoscale alignment of the suspension along the flow direction. Panel D illustrates \emph{in situ} polarized 
light imaging of shear-induced mesoscale alignment in the \emph{xz} plane of a CNC suspension as function of shear 
rate \(\dot\gamma\), with the same flow geometry sector visualized clockwise with increasing \(\dot\gamma\). 
The colored patterns contain information about mesoscale ordering of the CNC suspension.}    
\label{FIG:RheoSWAXS} 
\end{figure}

In parallel with the construction and commissioning of ForMAX, we have therefore also developed x-ray methodology to further 
expand the possibilities for multiscale and multimodal structural characterization during rheological and mechanical 
testing. Based on this development work we can, together with our sister beamline CoSAXS~\cite{kahnt21}, provide 
users with the following capability:
\begin{itemize}

\item Simultaneous rheological and SWAXS experiments, as exemplified in Figs.~\ref{FIG:RheoSWAXS}A and \ref{FIG:RheoSWAXS}C for a cellulose 
nanocrystal suspension subjected to laminar Couette flow in a concentric polycarbonate cup-bob geometry. Other geometries, including plate-plate 
geometry that allows simultaneous mesoscale structural characterization by polarized light imaging (see 
Fig.~\ref{FIG:RheoSWAXS}D), and environmental control are also available. Finally, we note that the MLM provides the prospect of supreme 
temporal resolution in such studies. 

\item We're addressing the need for more complex \emph{in situ} load experiments by developing combined dynamical 
mechanical analysis (DMA) and SWAXS in an atmosphere of controlled temperature and humidity, see 
Fig.~\ref{FIG:RheoSWAXS}B. Inspired by recent development of combined rheological and SR\(\mu\)CT 
experiments~\cite{dobson20}, we're currently expanding the DMA-SWAXS experiments towards multiscale structural 
characterization by introducing simultaneous SR\(\mu\)CT capability, using the rheometer in co-rotation mode as the 
tomographic rotation stage.
\end{itemize}
   
\section{Conclusions and outlook}

We have recently brought into operation ForMAX, that allows unique multiscale and multimodal structural characterization 
of hierarchical materials in the nm to mm range by combining SWAXS, scanning SWAXS imaging, and SR\(\mu\)CT (or any 
combination of these techniques) in a single experiment. Although we are still optimizing the beamline's performance, 
the initial benchmarking of the x-ray beam properties reported here demonstrates ForMAX's potential. 

A major aspect of ForMAX is the possibility to monitor multiscale structural evolution during material processing. 
Currently we are developing this possibility along two different paths. First, the very high photon density at ForMAX 
provides unprecedented possibilities for ultrafast full-field imaging. Second, we are working on dedicated sample 
environments that allow multiscale structural characterization during complex rheological or mechanical testing under 
controlled temperature and humidity, as exemplified above. We hope to make these developments available for general 
users in the near future. 

We thank the staff at the MAX IV Laboratory for all their support during the beamline project. The construction 
of ForMAX has been funded by the Knut and Alice Wallenberg  Foundation, while a large fraction of the operation costs 
are covered by Swedish industry via the Treesearch platform. 
Research conducted at MAX IV, a Swedish national user facility, is supported by the Swedish Research Council under contract 2018-07152, the 
Swedish Governmental Agency for Innovation Systems under contract 2018-04969, and Formas under contract 2019-02496.
We have received additional funding from the European Union's Horizon 2020 Framework Programme via the European Research Council 
(WIREDIRECT, grant agreement 801847; 3DX-FLASH, 948426; MUMOTT, 949301) and the Marie Sklodowska-Curie Actions (PSI-FELLOW-III-3i, grant 
agreement 884104), the Swedish Research Council (project grant number 2022-04192), the Swedish Foundation for Strategic 
Research (SSF grant ID17-0097), the Crafoord foundation, NanoLund, the 'FibRe - Competence Centre for Design for Circularity: 
Lignocellulose- based Thermoplastics' partly funded by the Swedish Innovation Agency VINNOVA (Grant Number 2019-00047), 
Chalmers Center for Advanced Neutron and X-ray scattering techniques, and Chalmers Area of Advance Materials Science.


\begin{table}
\caption{Main parameters of the MAX IV 3 GeV storage ring.} 
\label{TAB:MAXIV}
\begin{tabular}{ll}
\hline
Storage ring energy & 3 GeV\\ 
Circumference & 528 m\\ 
Beam current (operation November 2023) & 400 mA \\
Projected beam current & 500 mA \\
Electron beam emittance (\(x \times y \)) & \(326 \times 8\)~pm\(^2 \)rad\(^2 \) \\ 
Electron energy spread & \(7.7 \times 10^{-4}\) \\ 
Electron source size (\(x \times y \)) & \(54 \times 4~\mu\)m\(^2 \) \\ 
Electron source divergence (\(x \times y \)) & \(6 \times 2~\mu\)rad\(^2 \) \\ 
Top up  & Every 10 minutes \\ 
\hline
\end{tabular}
\end{table}

\begin{table}
\caption{Main components of the beamline.} 
\label{TAB:BL}
\begin{tabular}{ll}
\hline
Component & Distance from source [m] \\ \hline
Undulator & 0 \\ 
Front end movable mask & 19.5\\ 
White-beam slits & 23.9 \\
Double multilayer monochromator &  25.0 \\
Double crystal monochromator & 27.0 \\
Vertically focusing mirror & 30.2 \\
Horizontally focusing mirror & 31.0  \\
Monochromatic slits & 28.1, 32.3, 36.3, 41.5 - 41.8 \\
X-ray prism lens (placeholder) & 36.6 \\
Compound refractive lenses & 40.5 \\
Experimental table & 42.0 \\
Full-field microscope & 42.0 - 42.3\\
WAXS detector & 42.1 \\
SAXS detector & 42.8 - 49.6 \\ \hline
\end{tabular}
\end{table}

\begin{table}
\caption{Main parameters of the undulator.} 
\label{TAB:ID}
\begin{tabular}{ll}
\hline
Magnet material & NdFeB \\ 
Pole Material & Vanadium permedur \\
Energy range & 8-25 keV \\
Period length & 17 mm \\
Number of periods & 166 \\
Minimum magnetic gap & 4.5 mm \\
{\em K} value at minimum gap & 1.89 \\
Phase error &  \(\leq 2.5^\circ \) \\ 
Total power (projected 500 mA ring current) &  \(\approx 11.5\)~kW \\ \hline
\end{tabular}
\end{table}

\begin{table}
\caption{Hybrid photon-counting pixel detectors available at the beamline.} 
\label{TAB:detectors}
\begin{tabular}{lll}
\hline
& Dectris & X-Spectrum \\ 
 & EIGER2 X 4M & Lambda 3M \\ \hline 
Number of pixels  & 4M & 3M \\ 
Sensor size & \(2068\times 2162\) pixels & \(4\times 516\times 1536\) pixels \\ 
Pixel size & \(75\times 75\)~\(\mu\)m\(^2\)& \(55\times 55\)~\(\mu\)m\(^2\)  \\ 
Sensor material & Si  & Si \\ 
Sensor thickness & 450~\(\mu\)m  & 320~\(\mu\)m \\ 
Dynamic range & 32~bit  & 24~bit \\ 
Maximum frame rate\footnote{Full dynamic range.} & 560~Hz  & 1000~Hz \\ 
Data storage  & Streaming  & Streaming \\ 
Speciality & Vacuum compatible  & 'Windmill' shaped \\ 
\hline
\end{tabular}
\end{table}

\begin{table}
\caption{sCMOS cameras available at the beamline.} 
\label{TAB:cameras}
\begin{tabular}{llll}
\hline
 & Hamamatsu & Andor & Photron \\ 
& ORCA Lightning & Zyla 5.5 & FASTCAM Nova S16\\ \hline 
Number of pixels  & 12M & 5.5M  & 1M \\ 
Sensor size & \(4608\times 2592\) pixels & \(2560\times 2160\) pixels & \(1024\times 1024\) pixels \\ 
Pixel size & \(5.5\times 5.5\)~\(\mu\)m\(^2\) &  \(6.5\times 6.5\)~\(\mu\)m\(^2\) &  \(20\times 20\)~\(\mu\)m\(^2\) \\ 
Maximum dynamic range & 16~bit & 16~bit & 12~bit\\ 
Maximum frame rate\footnote{Full frame} & 121~Hz (12 bit) & 100~Hz (12 bit) & 16~kHz (12 bit)\\ 
 & 30~Hz (16 bit) & 75~Hz (16 bit) & \\ 
Data storage & Streaming & Streaming & 128 Gb / 4 Tb\\ 
\hline
\end{tabular}
\end{table}

\end{document}